\documentclass[aps,pra,twocolumn,showpacs]{revtex4}
\usepackage{amsmath,amssymb,graphicx,bbold}
\usepackage{dcolumn}   % needed for some tables
\usepackage{bm}        % for math
\usepackage{color}
\usepackage[utf8]{inputenc} %pacote de acento e ç 

\providecommand{\abs}[1]{\left\lvert#1\right\rvert}

\providecommand{\ket}[1]{\lvert #1 \rangle}
\providecommand{\bigbra}[1]{\left\langle #1 \right\rvert}
\providecommand{\bigket}[1]{\left\lvert #1 \right\rangle}

\begin{document}

\title{Astigmatic tomography of orbital angular momentum superpositions}
\author{B. Pinheiro da Silva, D. S. Tasca, E. F. Galv\~ao, A. Z. Khoury}

\affiliation{
Instituto de F\'\i sica, Universidade Federal Fluminense,
24210-346 Niter\'oi - RJ, Brasil}
\date{\today}

\begin{abstract}
We use astigmatic transformations to characterize two-dimensional superpositions of Orbital Angular Momentum (OAM) states in laser beams. We propose two methods for doing this, both relying only on astigmatic transformations, viewed as rotations on the Poincar\'e sphere, followed by imaging. These methods can be used as a tomographic tool for communication protocols based on optical vortices. 

%The astigmatic effect is used to characterize optical qubits encoded on the orbital angular momentum of a paraxial beam. The method is naturally described in terms of rotations in the Poincar\'e sphere and can be used as a tomographic tool for quantum information protocols based on optical vortices. 
\end{abstract}
%\ocis{270.5585,270.5565}
%\pacs{42.50.Dv}
%\vskip2pc 

\maketitle

\section{introduction}
\label{intro}

The orbital angular momentum (OAM) of light has been used for numerous applications, including optical tweezing \cite{tweezer1,tweezer2}
and communication protocols \cite{cryptouff,cryptolorenzo,tamburini,krenn}. 
The ability to generate and detect OAM beams has become imperative for such applications \cite{oam25}. The 
demonstration of OAM entanglement in photon pairs generated by spontaneous parametric down conversion has brought OAM into the 
experimental toolbox of quantum information \cite{zeilinger}. Nowadays, several methods for generating, transforming and detecting OAM beams can 
be found in the literature \cite{oam-masurement1,oam-masurement2,oam-sorter1,Schlederer16,oam-sorter2,oam-electron}. 
The astigmatic mode conversion is a powerful tool and has been widely used for implementing unitary transformations in the OAM degree 
of freedom. 
The astigmatic effect can be introduced either by cylindrical lenses \cite{woerdman,kimel,cnotuff} or by tilted spherical lenses \cite{tilted}.  

The formal equivalence between first order paraxial modes and light polarization naturally leads to a Poincar\'e sphere 
representation of these modes and their use as an additional qubit for quantum information encoding \cite{poincare}. 
In this way, two qubits can be encoded in a single photon, one on its polarization state and the other on its OAM structure 
\cite{topoluff,quplate,cardano,leo1,milione,aiello2,structured-waves}. 
One central issue in quantum information processing is the characterization of a qubit source, determining the quantum state it produces. 
The most common approach allowing the tomographic reconstruction of the transverse spatial state of paraxial light fields
employs projective measurements based on the phase-flattening of the field wavefront \cite{Giovannini13,Qassim14,Rathore17}.
More recently, wavefront reconstruction methods exploring digital holography \cite{DErrico17,Knudsen17} have also been used to characterize the transverse spatial state in terms of its OAM constituents.

% computer generated holograms \cite{Schulze12} or

%Despite the numerous methods available for OAM measurements, determination of arbitrary superposition states is a difficult task even for OAM qubits. 

In this work we exploit the astigmatic effect to measure arbitrary superpositions of OAM superpositions. 
%This can be useful as a tomographic tool for communication protocols based on optical OAM.
%Our method is based on the Poincar\'e sphere representation of OAM modes \dan{of first order} and can be geometrically interpreted in terms of rotations in the sphere. 
We provide two distinct experimental demonstrations of the use of astigmatic optical elements to obtain the spherical coordinates of first order OAM modes in the Poincar\'e sphere. Our methods can be geometrically interpreted in terms of rotations in the sphere and require at most three images of the astigmatically-transformed beam under investigation. We demonstrate these methods with beams prepared in two-dimensional superpositions of OAM states with topological charges $\pm 1\,$, but their extension to two-dimensional superpositions of higher topological charges $\pm l$ $(l>1)$ is straightforward. 

The paper is structured as follows. In section \ref{astig} we briefly review the mathematical representation of astigmatic transformations. In section \ref{tilted} we describe a simple tomographic method based on the visibility maximization of the astigmatically transformed beam. This visibility is maximized over a continuous parameter (an angle) of the astigmatic transformations. Our experimental scheme and supporting experimental results are presented in section \ref{preparation}. In section \ref{geo} we describe a second tomographic method utilizing a fixed number ($\leqslant$ 2) of astigmatic transformations and test it with our experimental data. Conclusions are presented in section \ref{conclusion}.

\section{Review: astigmatic transformations on the Poincar\'e sphere}
\label{astig}

In this section we review how optical vortices with Orbital Angular Momentum (OAM) topological charge $\pm1$ can be pictured on the Poincar\'e sphere. We also describe astigmatic transformations on these states as rotations of the Poincar\'e sphere. Later, in sections \ref{tilted} and \ref{geo}, we describe two different ways to characterise general states on the Poincar\'e sphere using only astigmatic transformations and the orientation of the resulting HG modes, obtained by imaging.

Pure vortex beams are characterized by a rotationally symmetric intensity distribution around the propagation axis accompanied 
by the azimuthal variation of the phase. This makes it difficult to measure the topological charge of an optical vortex with 
intensity only measurements. In general one must resort to interferometry in order to resolve the azimuthal phase variation \cite{DErrico17,Knudsen17}. 
However, the astigmatic mode conversion can transform pure Laguerre-Gaussian (LG) beams into Hermite-Gaussian (HG) intensity 
distributions that can be directly identified through intensity measurements. For example, the astigmatic effect provided by a 
spherical tilted lens has been used as a tool for topological charge measurements \cite{tilted}. 
Rather than simply measuring the topological charge, the ability to resolve OAM superpositions is a much more difficult task. 
Resolution of the superposition constituents, their weights and relative phases cannot be easily measured at once. 

\begin{figure}
	\centering
	\includegraphics[scale=0.8]{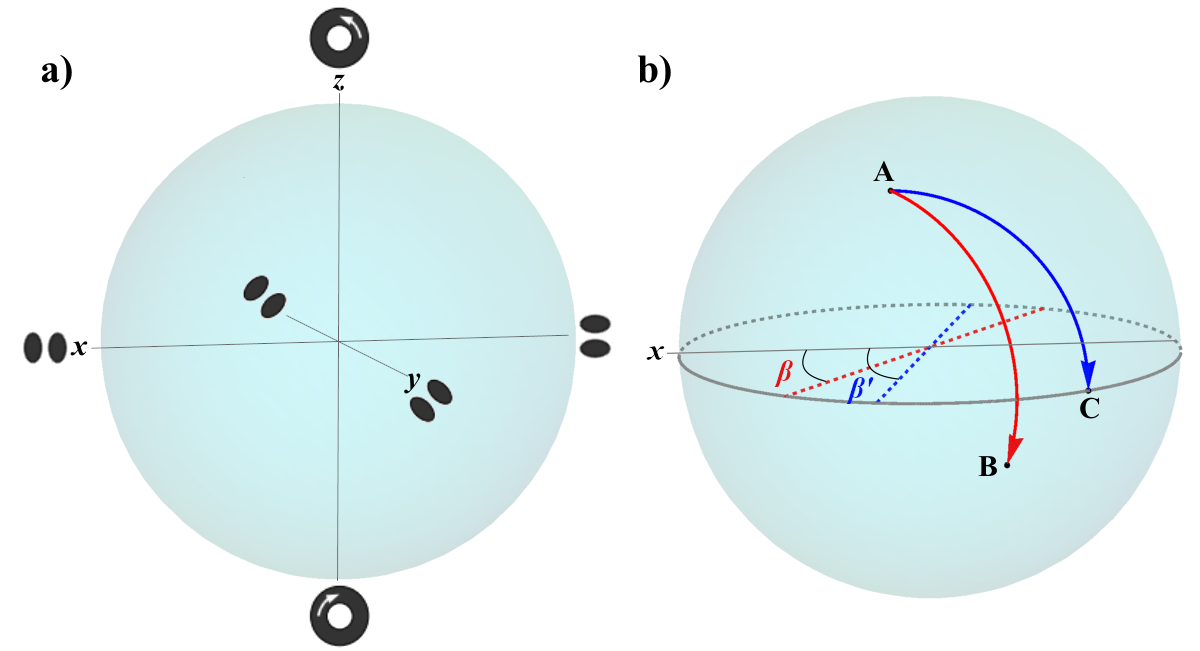}
	\caption{a) Poincar\'e sphere epresentation of the two-dimensional space of Orbital Angular Momentum topological charge $\pm 1$. b) The astigmatic transformations we apply correspond to rotations on the Poincar\'e sphere, by angle $-\pi/2$ around an axis on the equator. Here we picture one such rotation which does not map the initial point to the equator (red), and one which does (blue).}
	\label{fig:poincarerotation}
\end{figure}

However, if we restrict the OAM degree of freedom to a two-dimensional subspace, it is possible to use astigmatic 
transformations to resolve arbitrary superpositions with intensity only measurements. In this work we manipulate arbitrary 
superpositions of two OAM beams with opposite topological charges $\pm 1\,$,
\begin{eqnarray} \label{InputState}
\psi_{\theta,\phi}(\mathbf{r}) &=& \cos\left(\frac{\theta}{2}\right)\,LG_{+} (\mathbf{r}) 
+ e^{i\phi} \sin\left(\frac{\theta}{2}\right)\,LG_{-} (\mathbf{r})\,,
\nonumber\\
\end{eqnarray}
where $LG_{\pm}$ is the Laguerre-Gaussian mode function with topological charge $\pm 1\,$, $\mathbf{r}$ is the transverse position and $\theta$ and $\phi$ are, 
respectively, the sagittal and azimuthal angles in the Poincar\'e sphere representation.  This general superposition can be formally represented as the column vector
\begin{eqnarray}
\ket{\theta,\phi} &=& 
\left[
\begin{matrix}
\cos\frac{\theta}{2}\\
e^{i\phi} \sin\frac{\theta}{2}\\
\end{matrix}
\right]\;.
\label{sup}
\end{eqnarray}
This two-dimensional complex vector space is analogous to the Hilbert space of a two-level system, a qubit. Accordingly, there is a one-to-one correspondence between each state (\ref{sup}) and points on the surface of the Poincar\'e sphere (see Fig. \ref{fig:poincarerotation}-a). 
 
The action of an astigmatic mode converter is conveniently represented using this matrix formulation. An LG-HG mode converter rotated by $\beta/2$ with respect to the horizontal introduces a $\pi/2$ retardation between HG modes oriented at $\beta/2$ and $(\beta + \pi)/2\,$. This is naturally represented by 
\begin{eqnarray}
MC(\beta) &=& \bigket{\frac{\pi}{2},\beta} \bigbra{\frac{\pi}{2},\beta} - i \bigket{\frac{\pi}{2},\beta+\pi}\bigbra{\frac{\pi}{2},\beta+\pi}
\nonumber\\
&=& \frac{e^{i\pi/4}}{\sqrt{2}}
\left[
\begin{matrix}
1 & i\,e^{-i\beta}\\
i\,e^{i\beta} & 1\\
\end{matrix}
\right]\;.
\label{matrixastig}
\end{eqnarray}
This transformation is a good approximation to the one implemented either by a LG-HG mode converter, or by a tilted spherical lens, as proposed in \cite{tilted}. Geometrically, this is a rotation by an angle $-\pi/2$ around an axis on the equator of the Poincar\'e sphere, situated at angle $\beta$ with the $x$ axis (see Fig. \ref{fig:poincarerotation}-b).

In the next section we describe our first method to characterize arbitrary superpositions of first order OAM modes on the Poincar\'e sphere.

\section{Method I: Hermite-Gaussian mode visibility}
\label{tilted}

As we have seen in the previous section, an astigmatic LG-HG mode converter rotates states by $-\pi/2$ radians around an axis in the equator of the Poincar\'e sphere. The orientation of the rotation axis with respect to the $x$ axis is given by $\beta$, where $\beta/2$ is the physical angle of the LG-HG mode converter with respect to the horizontal.

It is a geometrical fact that any point (A) on a sphere is mapped to some point (C) on the equator by a $\pi/2$ rotation around a suitable axis on the equator (see the blue arc in Fig. \ref{fig:poincarerotation}-b). Thus, for any state on the Poincar\'e sphere, there must be a LG-HG mode converter orientation that takes the state onto the equator. Points on the equator correspond to rotated HG modes which can be easily identified from intensity measurements (images) only.

After passing through the mode converter, the general superposition (\ref{sup}) is transformed into 
\begin{eqnarray}
MC(\beta)\ket{\theta,\phi} &=& 
\frac{e^{i\pi/4}}{\sqrt{2}}
\left[
\begin{matrix}
\cos\frac{\theta}{2} + i\,e^{i(\phi-\beta)} \sin\frac{\theta}{2}\\
\\
i\,e^{i\beta} \cos\frac{\theta}{2} + \,e^{i\phi} \sin\frac{\theta}{2}\\
\end{matrix}
\right]\;.
\label{mcsup}
\end{eqnarray}
Our first method to determine the transverse mode state involves rotating the mode converter until we find the appropriate angle $\beta$ for which the general mode (\ref{sup}) can be recognised by imaging as a rotated HG mode. In this case, the two column entries must have the same absolute values
\begin{eqnarray}
\abs{\cos\frac{\theta}{2} + i\,e^{i(\phi-\beta)} \sin\frac{\theta}{2}}
&=&
\abs{i\,e^{i\beta} \cos\frac{\theta}{2} + \,e^{i\phi} \sin\frac{\theta}{2}}\;.
\nonumber\\
\label{equalmod}
\end{eqnarray}
Of course, this condition is naturally fulfilled for $\theta = 0$ or $\pi\,$, regardless 
to the value of $\beta$ (pure LG modes are transformed into HG modes for any orientation of 
the mode converter). For arbitrary $\theta\,$, condition (\ref{equalmod}) brings us to 
\begin{eqnarray}
\beta_{MC} &=& \phi\;.
\label{betamc}
\end{eqnarray}
The orientation of the resulting HG mode with respect to the horizontal is given by 
\begin{eqnarray}
\alpha_{HG} &=& \frac{1}{2}\arg\left(\frac{i\,e^{i\beta} \cos\frac{\theta}{2} + \,e^{i\phi} \sin\frac{\theta}{2}}
{\cos\frac{\theta}{2} + i\,e^{i(\phi-\beta)} \sin\frac{\theta}{2}}\right) 
\nonumber\\
&=& \frac{\phi-\theta}{2} + 45^\circ \,.
\label{alphahg}
\end{eqnarray}
Therefore, an arbitrary superposition can be characterized by two angles. The first is the mode converter orientation $\beta_{MC}/2\,$ with respect to the horizontal, which takes the superposition into a rotated HG mode. The second angle is the resulting orientation $\alpha_{HG}\,$ of the HG mode found in this way.

\subsection{Experimental implementation} 
\label{preparation}

In order to apply the result of the previous section, we start by preparing various superpositions of $LG_{\pm}$ 
modes, sending them through an astigmatic analyzer. The superposition parameters 
$(\theta,\phi)$ are then inferred from the measured values of $\beta_{MC}$ and $\alpha_{HG}$ according to 
Eqs. (\ref{betamc}) and (\ref{alphahg}).

\begin{figure}
	\centering
	\includegraphics[scale=0.23]{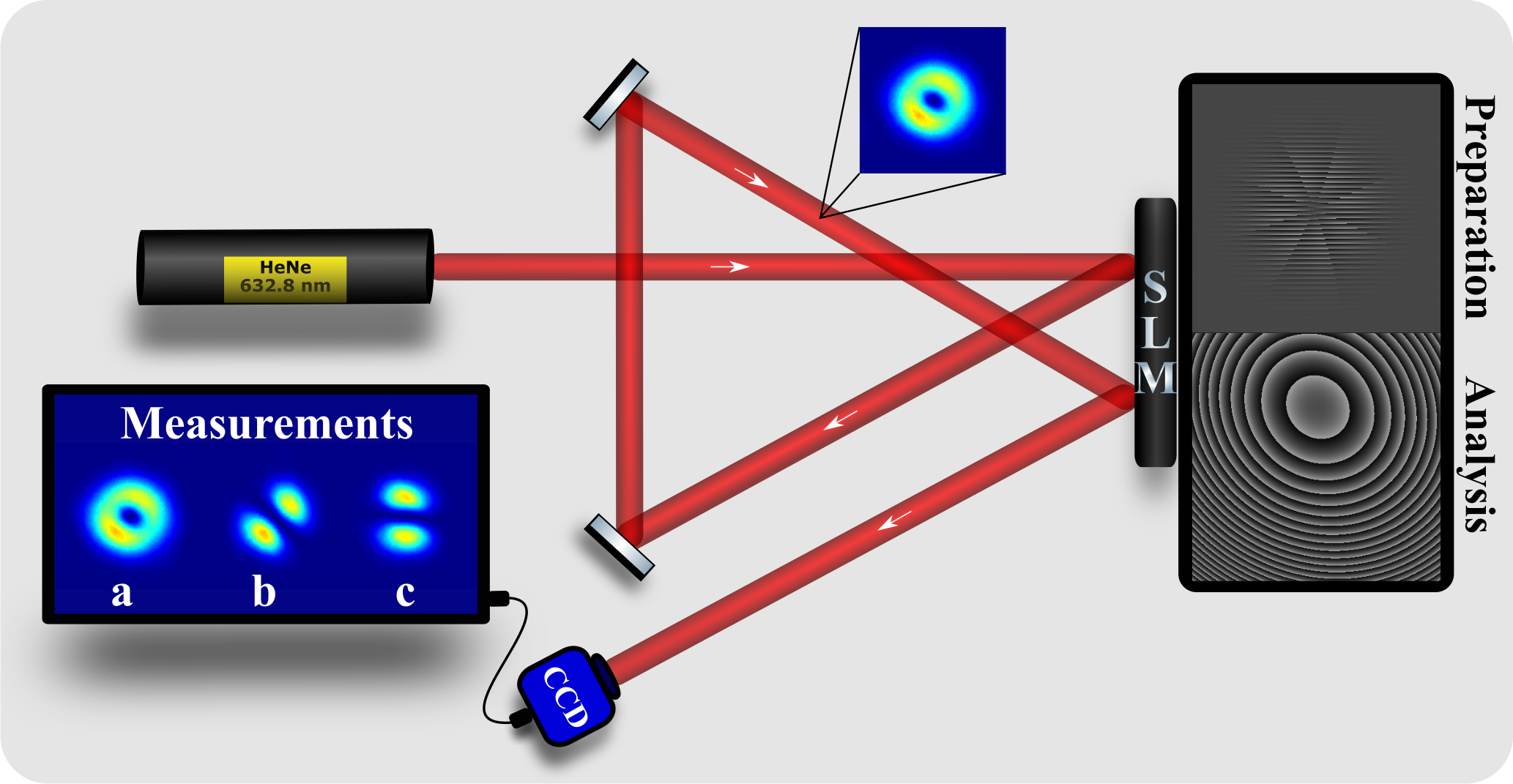}
	\caption{Experimental setup. The screen of the SLM was split in two sides, one for the preparation of the mode and the other to emulate the tilted lens. The screen connected to the CCD shows the three images used in the method II, these images are not captured simultaneously.}
	\label{fig:setup}
\end{figure}

The experimental setup is depicted in Fig. \ref{fig:setup}. The Gaussian beam 
from a He-Ne laser ($\lambda = 633nm$) is sent to one half of a reflective spatial light modulator (SLM) panel, 
programmed to create a controllable OAM state $\ket{\theta,\phi}\,$. Then, the astigmatic analyzer is 
implemented in the other half of the SLM, programmed with a phase modulation that emulates the transmission 
function of a tilted lens \cite{tilted}. A charge coupled device (CCD) camera positioned close to the focal plane of this tilted lens is used
to acquire the image of the transformed beam.
%The image produced close to the focal plane is acquired by a charge coupled device (CCD) camera. 
Two parameters are set by the analyzer: the tilt angle $\xi\,$, which remains fixed 
once calibrated, and the angle $\beta/2$, as shown in Fig. \ref{fig:lente3d}.
The astigmatic phase profile written on the SLM is provided in appendix \ref{SLMPhase}.

\begin{figure}[b!]
	\centering
	\includegraphics[scale=0.4]{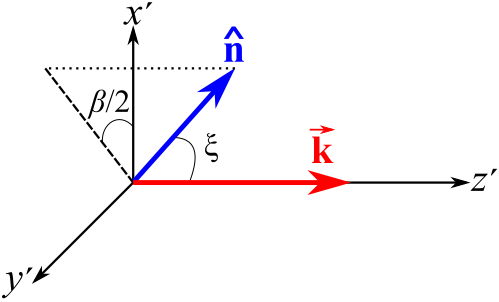}
	\caption{Tilted lens orientation. A spherical lens (axis $\mathbf{\hat{n}}$) implements the astigmatic mode conversion given by Eq. (\ref{matrixastig}). The tilt angle $\xi$ is determined experimentally, so that a rotation of $-\pi/2$ radians on the Poincar\'e sphere is achieved. The axis of the rotation on the Poincar\'e sphere is on its equator, at an angle $\beta$ with the $x$ axis, determined by the angle of the projection of $\mathbf{\hat{n}}$ with the transverse plane, as in the figure.}
	\label{fig:lente3d}
\end{figure}

First, the tilt angle is calibrated by sending a pure LG beam to the SLM analyzer and verifying the correct formation of a pure HG profile close to the focal plane, as described in Ref. \cite{tilted}. This calibration is independent of the lens axis orientation, which is initially placed at the horizontal plane. Once calibration 
is achieved, the analyzer is ready to receive the sample OAM states. The analysis is performed by keeping fixed the tilt angle and turning the lens axis $\mathbf{\hat{n}}$ around the beam propagation direction until a pure HG profile is found in the observation plane. At this moment, we take note of two angles: the orientation $\beta_{MC}/2$ of the transverse projection of the lens axis $\mathbf{\hat{n}}$ with respect to the $x'$ axis (see Fig. \ref{fig:lente3d}), and the orientation $\alpha_{HG}$ of the line connecting the maxima of the resulting HG mode, with respect to the horizontal (see appendix \ref{algorithm}). According to Eqs. (\ref{betamc}) and (\ref{alphahg}), they determine the Poincar\'e sphere coordinates $(\theta,\phi)$ of the sample state.

\begin{figure}
	\centering
	\includegraphics[scale=0.25]{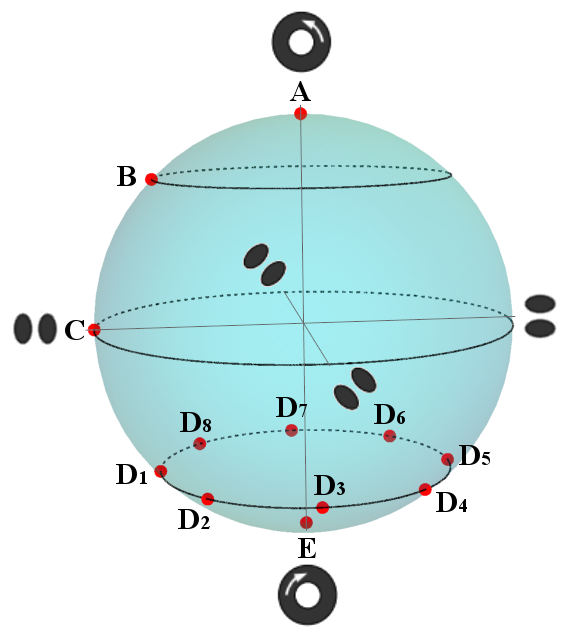}
	\caption{Poincar\'e sphere representation of the sample modes used for testing the astigmatic analyzer.}
	\label{fig:pontos1}
\end{figure}

\begin{figure}
	\centering
	\includegraphics[scale=0.22]{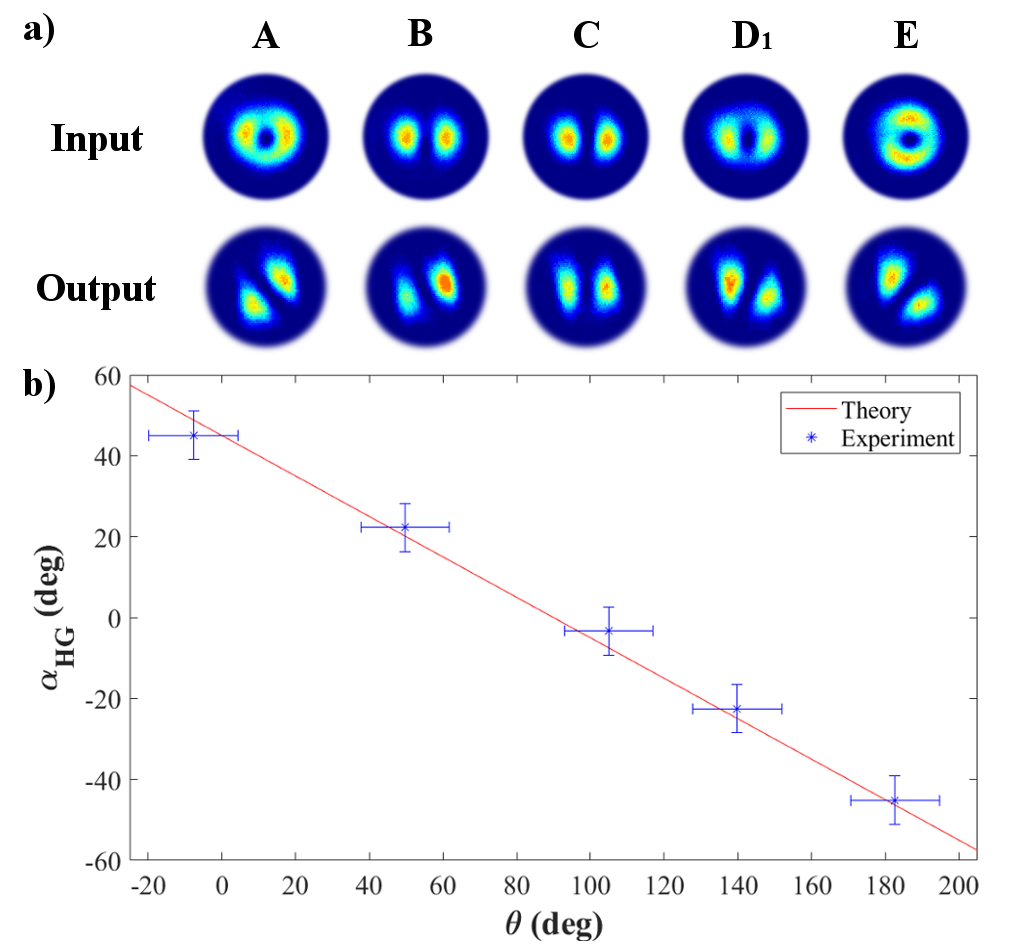}
	\caption{a) Images of the input modes and the resulting outputs from the astigmatic analyzer for the sequence A-E.
		b) Graphical representation of the measured orientations $\alpha_{HG}$ versus $\theta\,$. 
	Blue dots are the experimental results and red solid line is the theoretical prediction given by Eq. (\ref{alphahg}).}
	\label{fig:grafico1}
\end{figure}

The astigmatic mode analysis has been tested with two sequences of sample states, as represented in Fig. \ref{fig:pontos1}. The first sequence (A-E) covers the path along the meridian $\phi=0$ and illustrates the relationship between $\alpha_{HG}$ and 
$\theta\,$. In Fig. \ref{fig:grafico1}a we show the input modes produced by the preparation screen and the resulting HG profiles at the output of the analyzer. The graphical representation of $\alpha_{HG}$ versus $\theta$ is presented in Fig. \ref{fig:grafico1}b. 


\begin{figure}
	\centering
	\includegraphics[scale=0.22]{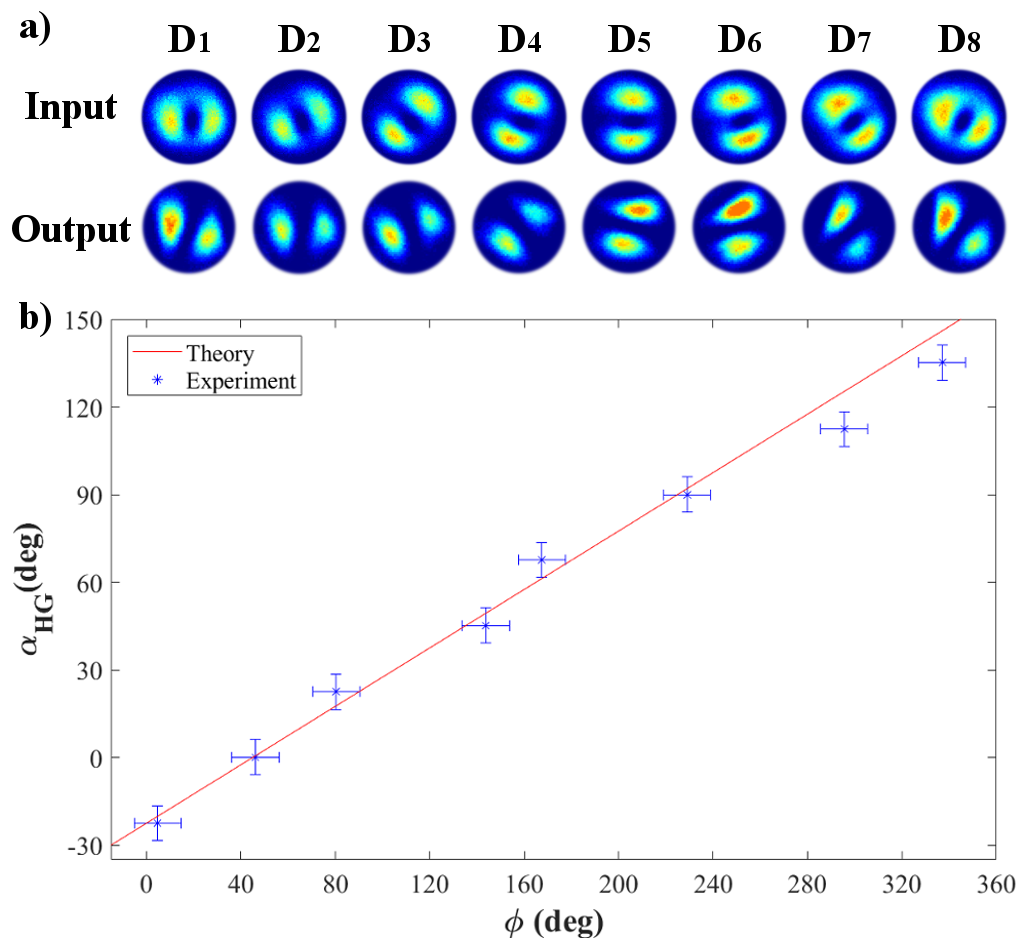}
	\caption{a) Images of the input modes and the resulting outputs from the astigmatic analyzer for the sequence D1-D8.
		b) Graphical representation of the measured orientations $\alpha_{HG}$ versus $\phi\,$. 
		Blue dots are the experimental results and red solid line is the theoretical prediction given by Eq. (\ref{alphahg}).}
	\label{fig:grafico2}
\end{figure}

The second sequence (D1-D8) covers the path along the polar circle $\theta=135^\circ$ and illustrates the relationship between $\alpha_{HG}$ and $\phi\,$. In Fig. \ref{fig:grafico2}a we show the input modes produced by the preparation screen and the resulting HG profiles at the output of the analyzer. The graphical representation of $\alpha_{HG}$ versus $\phi$ is presented in Fig. \ref{fig:grafico2}b. 

The graphs presented in Figs. \ref{fig:grafico1} and \ref{fig:grafico2} show a good agreement between the experimental results and the theoretical expression (\ref{alphahg}). Errors due to the experimental determination of $\beta_{MC}$ and $\alpha_{HG}$ are propagated and expressed as error bars. The average errors achieved with this method are $\Delta \theta=12^\circ$ and $\Delta \phi=10^\circ$. These errors are relatively large. In the next section we describe a second method that also uses astigmatic transformations and imaging to obtain a more precise determination of the transverse mode state.

\section{Method II: Poincar\'e sphere positioning}
\label{geo}

Of course, pure intensity measurements alone cannot provide full information about arbitrary OAM superpositions. However, partial information can be obtained by measuring the intensity distribution and the orientation $\alpha$ of the line connecting the two intensity maxima (with respect to the horizontal). This gives the azimuthal coordinate $\phi = 2\alpha$ of the mode superposition in question. Therefore, from pure intensity measurements we determine the meridian line, in the Poincar\'e sphere, passing through the point representing the state. Note, however, that the image of states too close to the poles on the $z$ axis will present low visibility of the maxima, as the states are close to LG states, for which the intensity distribution is symmetrical around the center. We should treat these polar regions as blind spots for our longitude determination; they will be discriminated by other measurements, as we now describe.
 
As we have seen, the astigmatic mode conversion can be geometrically interpreted as a $-\pi/2$ rotation of the Poincar\'e sphere around the axis connecting the mode conversion eigenstates. For example, a horizontal mode converter (eigenstates $\ket{\pi/2,0}$ and $\ket{\pi/2,\pi}$) rotates states by $-\pi/2$ radians around the $x$ axis, mapping all states on the meridians $\phi = 0$ and $\phi = \pi$ onto the equator.

\begin{figure}
	\centering
	\includegraphics[scale=0.22]{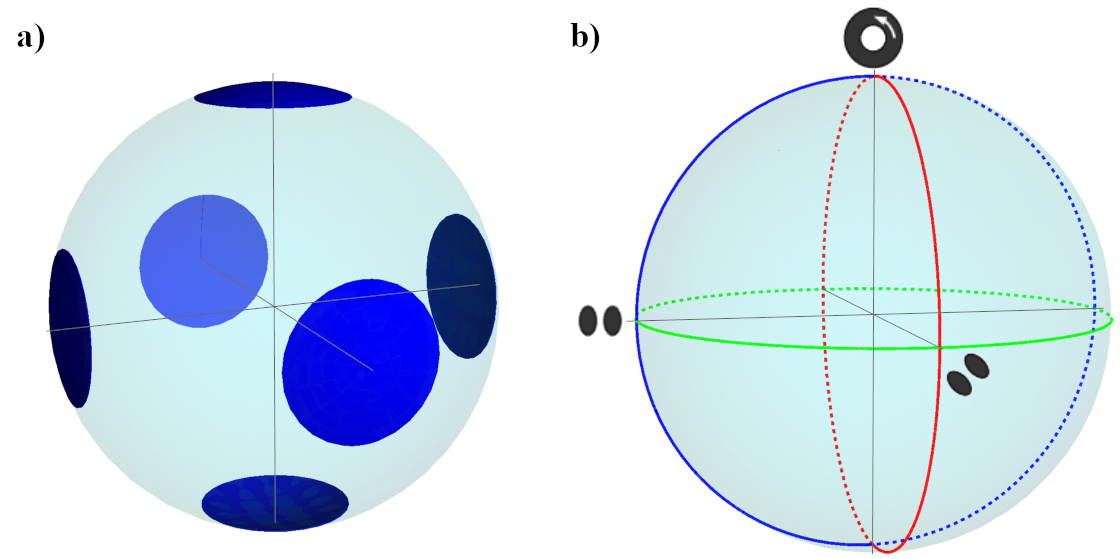}
	\caption{a) The spherical caps in blue, around the $x,y$ and $z$ axes, represent states for which one of the three images will have a mode with low visibility. Due to this, we will call theses caps polar blind-spots. States in each cap will be determined from the two images which has modes with good visibility.	b) States near the three pictured great circles will result in an intersection between halves of great circles which corresponds to a very narrow triangle (see Fig. \ref{fig:spheretriang}-a). These will be obtained as the mid-point of the shortest triangle side (see Fig. \ref{fig:spheretriang}-a).}
	\label{fig:calotas}
\end{figure}

Our strategy for finding the position of states on the Poincar\'e sphere requires determining  the orientation $\alpha$ with respect to the horizontal of the two maxima in three images for each state we would like to characterise. The first is the direct image without mode conversion ($\alpha_{NL}$); the second ($\alpha_{0^\circ}$), after a transformation corresponding to a horizontal LG-HG mode converter ($\beta/2=0$); an the third ($\alpha_{45^\circ}$), after a mode converter oriented at $\beta/2=45^\circ$. As can be seen in Fig. \ref{fig:calotas}-a, each measurement will have low visibility for states in one of the six polar blind spots.

\begin{figure}
	\centering
	\includegraphics[scale=0.3]{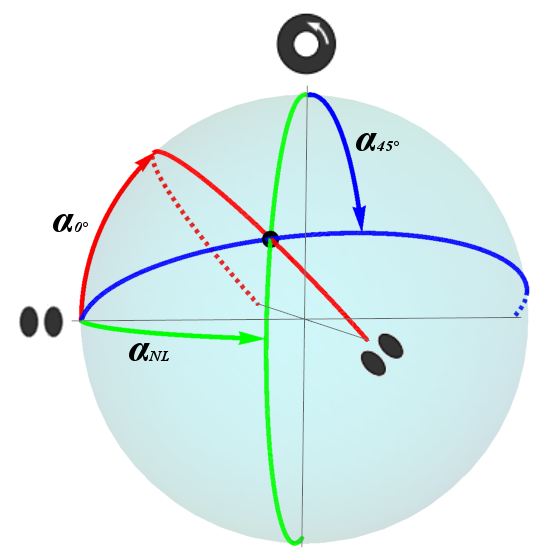}
	\caption{Poincar\'e representation of the three-measurement scheme. Characterisation of the angle that the line connecting the maxima makes with the horizontal gives information about the $z$- axis longitude (green). By previously rotating the Poincar\'e sphere via two astigmatic transformations, the same imaging technique gives us the $x$-axis longitude (blue) and $y$-axis longitude (red). In an idealized measurement, the three longitudes would intersect at the point representing the state.}
	\label{fig:gc}
\end{figure}

Ideally, this gives the input state as the intersection between three halves of great circles intersecting the $x, y$ and $z$ axes, as depicted in Fig. \ref{fig:gc}. Note that states on the $xy, xz, yz$ planes, i.e. on precisely the three great circles in Fig. \ref{fig:calotas}-b, are particular with respect to this characterization. This is because they sit on an overlap between two halves of great circles, and this overlap has a length of a quarter of a great circle. This means states will fail to be characterised as the intersection of two halves of great circles, hence the need for three images. Using the three images also guarantees that the polar, low-visibility ``blind spots" can always be avoided, as each of the six polar spots in Fig. \ref{fig:calotas}-a is only blind to a single measurement; in these cases, the two remaining measurements provide an intersection of two halves of a great circle, which determine the state's position.
\subsection{Experimental implementation}

The experimental scheme is as described in section \ref{preparation}, please see Fig. \ref{fig:setup}. We experimentally tested the method with the twenty six sample states represented by the red dots in Fig. \ref{fig:pontos2}. Three orientations ($\alpha_{NL},\alpha_{0^\circ},\alpha_{45^\circ}$) of the maxima with respect to the horizontal were measured for each sample state and  the state's parameters were estimated from the intersection between the three half great-circles inferred from each measurement. Due to experimental imperfections, we did not obtain a single intersection, but three corners of a triangular region on the Poincar\'e sphere.

For each point, we first test whether we are in one of the six polar blind spots of Fig. \ref{fig:calotas}-a. This was done by checking the visibility of each of the three images (see appendix \ref{algorithm}). A straighforward theoretical calculation determines that a minimum threshold of $v \le 0.34$ determines a polar cap of 0.38 steradians, as in Fig. \ref{fig:calotas}-a. In the cases when a point has a visibility below the threshold (i.e. is within a blind polar cap), we ignore the corresponding measurement/image, and the point's coordinates are determined by the intersection of the two halves of great circles corresponding to the other two measurements.

\begin{figure}
	\centering
	\includegraphics[scale=0.22]{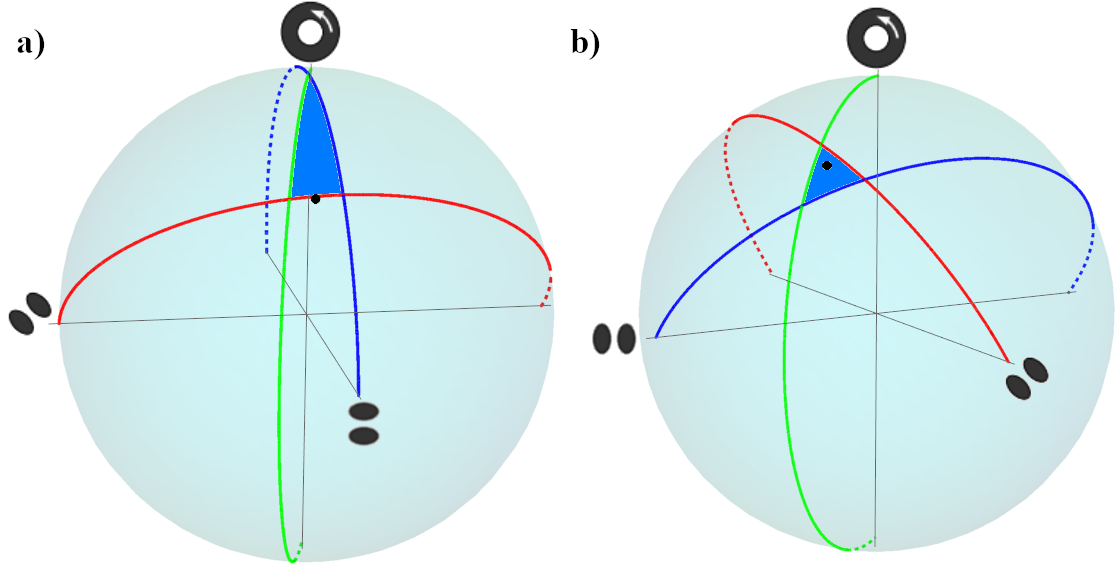}
	\caption{a) A state will be deemed near to the three problematic great circles of Fig. \ref{fig:calotas}-b when the three determined longitudes determine a narrow triangle (with shortest side smaller than half the next shortest side, in blue). In this case, we estimate the state as being at the half-point of the shortest side.	b) For generic states, far from the polar caps of Fig. \ref{fig:calotas}-a and the great circles of Fig. \ref{fig:calotas}-b, the state is deemed to be at the center of the spherical triangle (in blue).}
	\label{fig:spheretriang}
\end{figure}

\begin{figure}
	\centering
	\includegraphics[scale=0.3]{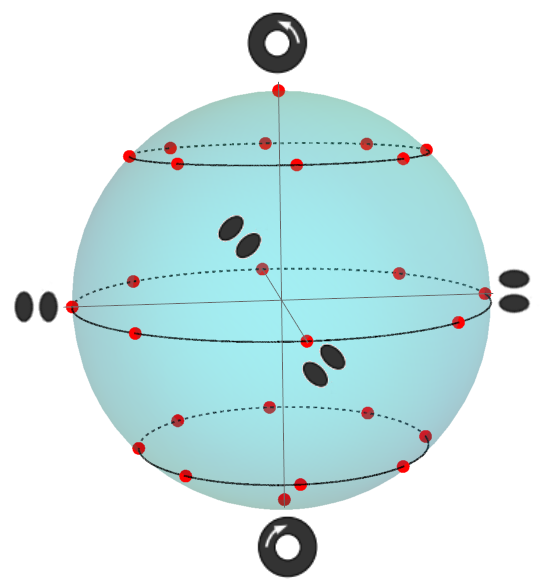}
	\caption{Poincar\'e representation of the sample states used to test the three-measurement scheme.}
	\label{fig:pontos2}
\end{figure}

Next, we experimentally evaluate whether the state is on, or near, the problematic great circles whose axes are the $x, y$ and $z$ axes (see Fig. \ref{fig:calotas}-b). This was done by comparing the sides of the triangular region determined by the three halves of great circles (see Fig. 9): if the shortest triangle side was shorter than half of the next shortest side, we considered the state was too close to one of the problematic great circles. In these cases, we estimated the state as lying on the mid-point of the shortest triangle side (see Fig. \ref{fig:spheretriang}-a). The distance between two vertices of the triangle can be obtained from their scalar product. Given that the vertices are on the surface of the Poincaré sphere with unit radius, the distance is simply given by 

\begin{equation}
\label{dist}
\mathrm{Dist}_{ab} = \arccos(\textbf{a} \cdot \textbf{b}), 
\end{equation}
where $\textbf{a}$ and $\textbf{b}$ are vectors that point to the vertices of the triangle.   

Let us assume now that the state is generic, i.e. it is both far from the blind polar regions of Fig. \ref{fig:calotas}-a, and from the problematic great circles of Fig. \ref{fig:calotas}-b. Then we estimate the state's parameters to be at the center of the triangular region determined by the three intersections of halves of great circles (see Fig. \ref{fig:spheretriang}-b), with error bars given by the largest variations of the state parameters within the triangular region.

In table \ref{tabela} we show our experimental results. We also calculated the fidelity between the measured and the target states, which evaluates the overall efficiency of the preparation and analysis process. The average error in determination of the angles was $\Delta \theta=5^\circ$ and $\Delta \phi=2^\circ$, while the average fidelity was $0.992 \pm 0.003 $.

\begin{table}
\begin{tabular}{|c|c|c|c|c|c|}
	
	\hline 
      Point & $\theta_T$ & $\phi_T$ & $\theta_E \pm  \Delta \theta_E$ & $\phi_E \pm  \Delta \phi_E$ & Fidelity   \\ 
	\hline 
	1 & 0\textordmasculine & 0\textordmasculine & 4\textordmasculine $\pm$ 2\textordmasculine & 121\textordmasculine $\pm$ 0\textordmasculine & 0.999 $\pm$ 0.001    \\ 
	\hline 
	
	2 & 45\textordmasculine & 0\textordmasculine & 41\textordmasculine $\pm$ 2\textordmasculine & 359\textordmasculine $\pm$ 5\textordmasculine & 0.999 $\pm$ 0.001    \\ 
	\hline 
	3 & 90\textordmasculine & 0\textordmasculine & 88\textordmasculine $\pm$ 2\textordmasculine & 354\textordmasculine $\pm$ 1\textordmasculine & 0.997 $\pm$ 0.001    \\ 
	\hline
	 
	4 & 135\textordmasculine & 0\textordmasculine & 140\textordmasculine $\pm$ 2\textordmasculine & 346\textordmasculine $\pm$ 1\textordmasculine & 0.991 $\pm$ 0.002    \\ 
	\hline
	5 & 180\textordmasculine & 0\textordmasculine & 166\textordmasculine $\pm$ 2\textordmasculine & 270\textordmasculine $\pm$ 0\textordmasculine & 0.986 $\pm$ 0.004    \\ 
	\hline
	
	6 & 45\textordmasculine & 45\textordmasculine & 47\textordmasculine $\pm$ 10\textordmasculine & 46\textordmasculine $\pm$ 6\textordmasculine & 0.999 $\pm$ 0.003    \\ 
	\hline 
	7 & 90\textordmasculine & 45\textordmasculine & 94\textordmasculine $\pm$ 14\textordmasculine & 36\textordmasculine $\pm$ 1\textordmasculine & 0.993 $\pm$ 0.009    \\ 
	\hline 
	
	8 & 135\textordmasculine & 45\textordmasculine & 145\textordmasculine $\pm$ 2\textordmasculine & 28\textordmasculine $\pm$ 1\textordmasculine & 0.983 $\pm$ 0.003    \\ 
	\hline 
	9 & 45\textordmasculine & 90\textordmasculine & 52\textordmasculine $\pm$ 2\textordmasculine & 83\textordmasculine $\pm$ 4\textordmasculine & 0.994 $\pm$ 0.003    \\ 
	\hline 
	
	10 & 90\textordmasculine & 90\textordmasculine & 102\textordmasculine $\pm$ 2\textordmasculine & 83\textordmasculine $\pm$ 1\textordmasculine & 0.985 $\pm$ 0.003    \\ 
	\hline 
	11 & 135\textordmasculine & 90\textordmasculine & 152\textordmasculine $\pm$ 2\textordmasculine & 91\textordmasculine $\pm$ 4\textordmasculine & 0.979 $\pm$ 0.005    \\ 
	\hline 
	
	12 & 45\textordmasculine & 135\textordmasculine & 44\textordmasculine $\pm$ 10\textordmasculine & 125\textordmasculine $\pm$ 5\textordmasculine & 0.997 $\pm$ 0.004    \\ 
	\hline 
	13 & 90\textordmasculine & 135\textordmasculine & 89\textordmasculine $\pm$ 7\textordmasculine & 137\textordmasculine $\pm$ 1\textordmasculine & 0.999 $\pm$ 0.002    \\ 
	\hline
	
	14 & 135\textordmasculine & 135\textordmasculine & 145\textordmasculine $\pm$ 2\textordmasculine & 158\textordmasculine $\pm$ 4\textordmasculine & 0.976 $\pm$ 0.007    \\ 
	\hline
	15 & 45\textordmasculine & 180\textordmasculine & 49\textordmasculine $\pm$ 2\textordmasculine & 171\textordmasculine $\pm$ 8\textordmasculine & 0.996 $\pm$ 0.006    \\ 
	\hline
	
	16 & 90\textordmasculine & 180\textordmasculine & 96\textordmasculine $\pm$ 2\textordmasculine & 186\textordmasculine $\pm$ 1\textordmasculine & 0.995 $\pm$ 0.002    \\ 
	\hline
	17 & 135\textordmasculine & 180\textordmasculine & 137\textordmasculine $\pm$ 3\textordmasculine & 210\textordmasculine $\pm$ 3\textordmasculine & 0.967 $\pm$ 0.007    \\ 
	\hline
	
	18 & 45\textordmasculine & 225\textordmasculine & 48\textordmasculine $\pm$ 3\textordmasculine & 213\textordmasculine $\pm$ 1\textordmasculine & 0.993 $\pm$ 0.002    \\ 
	\hline
	19 & 90\textordmasculine & 225\textordmasculine & 89\textordmasculine $\pm$ 12\textordmasculine & 230\textordmasculine $\pm$ 1\textordmasculine & 0.998 $\pm$ 0.003    \\ 
	\hline
	
	20 & 135\textordmasculine & 225\textordmasculine & 125\textordmasculine $\pm$ 2\textordmasculine & 244\textordmasculine $\pm$ 1\textordmasculine & 0.977 $\pm$ 0.004    \\ 
	\hline
	21 & 45\textordmasculine & 270\textordmasculine & 42\textordmasculine $\pm$ 2\textordmasculine & 259\textordmasculine $\pm$ 1\textordmasculine & 0.995 $\pm$ 0.001    \\ 
	\hline
	
	22 & 90\textordmasculine & 270\textordmasculine & 89\textordmasculine $\pm$ 2\textordmasculine & 272\textordmasculine $\pm$ 1\textordmasculine & 0.999 $\pm$ 0.001    \\ 
	\hline
	23 & 135\textordmasculine & 270\textordmasculine & 129\textordmasculine $\pm$ 2\textordmasculine & 273\textordmasculine $\pm$ 7\textordmasculine & 0.997 $\pm$ 0.002    \\ 
	\hline
	
	24 & 45\textordmasculine & 315\textordmasculine & 42\textordmasculine $\pm$ 6\textordmasculine & 306\textordmasculine $\pm$ 3\textordmasculine & 0.996 $\pm$ 0.003    \\ 
	\hline
	25 & 90\textordmasculine & 315\textordmasculine & 87\textordmasculine $\pm$ 16\textordmasculine & 313\textordmasculine $\pm$ 1\textordmasculine & 0.999 $\pm$ 0.008    \\ 
	\hline
	
	26 & 135\textordmasculine & 315\textordmasculine & 138\textordmasculine $\pm$ 8\textordmasculine & 309\textordmasculine $\pm$ 4\textordmasculine & 0.998 $\pm$ 0.004  \\
	\hline
\end{tabular} 
\caption{Experimental results for the sample points shown in Fig. \ref{fig:pontos2}.}
	\label{tabela}
\end{table}

\section{Conclusion}
\label{conclusion}

We developed two tomographic methods for the characterization of binary OAM superpositions based on astigmatic unitary transformations. The Poincar\'e sphere coordinates of arbitrary OAM superpositions could be inferred from intensity measurements only. The methods were implemented to characterize sets of sample states, showing high fidelities between the measured and generated states. We demonstrated these measurements with arbitrary superpositions of topological charges $\pm 1\,$, but their extension to two-dimensional superpositions of higher topological charges is straightforward. These methods can be useful for characterizing OAM sources in communication protocols.

\section*{Acknowledgments}
Funding was provided by 
Conselho Nacional de Desenvolvimento Tecnol\'ogico (CNPq), 
Coordena\c c\~{a}o de Aperfei\c coamento de 
Pessoal de N\'\i vel Superior (CAPES), Funda\c c\~{a}o de Amparo \`{a} 
Pesquisa do Estado do Rio de Janeiro (FAPERJ), and Instituto Nacional 
de Ci\^encia e Tecnologia de Informa\c c\~ao Qu\^antica (INCT-CNPq).

\appendix

\section{Astigmatic phase profile written on the SLM}
\label{SLMPhase}

The astigmatic phase profile written on the SLM panel and used in our experiment is that of a tilted, thin spherical lens. We set this phase profile to introduce astigmatism between rotated transverse coordinates defined as $x_\beta= \cos(\beta/2) \, x  + \sin(\beta/2) \, y$ and $y_\beta= - \sin(\beta/2) \, x + \cos(\beta/2)\,  y$, where $\beta/2$ is the angle of $x_\beta$ with respect to the horizontal axis $x=x_0$.
Upon propagation through this astigmatic phase mask, the input field $\psi_{\theta,\phi}$ given in Eq. \eqref{InputState} is transformed as
\begin{equation}
\psi_{\theta,\phi} \rightarrow \exp {\left[ -i \frac{k}{2f} \Phi(x_\beta,y_\beta)  \right] } \psi_{\theta,\phi},
\end{equation}
where $f=336$mm is the focal length of the programmed tilted-lens and $ \Phi(x_\beta,y_\beta)$ its phase profile, given as
\begin{equation}
\Phi(x_\beta,y_\beta) =    \sec(\xi) \, x_\beta^2 + \cos(\xi) \, y_\beta^2.
\end{equation}
The tilting angle $\xi$ used in our experiment is $27^\circ$.

\section{Algorithm to calculate mode inclination and visibility}
\label{algorithm}

\begin{figure}
	\centering
	\includegraphics[scale=0.6]{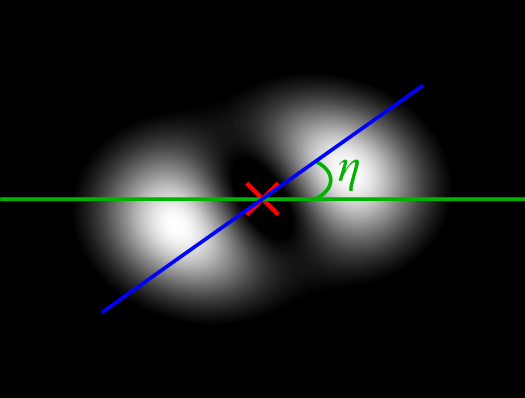}
	\caption{The red x is the image center of mass, the green line is the horizontal and the blue line is the one that we sum the intensity under it.   }
	\label{fig:varrer}
\end{figure}

\begin{figure}
	\centering
	\includegraphics[scale=0.6]{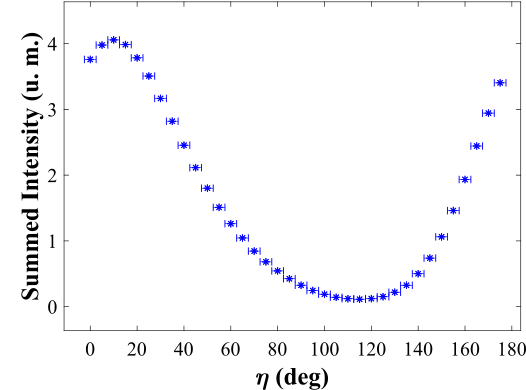}
	\caption{Graphical representation of the summed intensity versus $\eta$. The vertical error bar is very small and could not be represented within graphical precision. }
	\label{fig:grafico3}
\end{figure}

\begin{figure}[!h]
	\centering
	\includegraphics[scale=0.6]{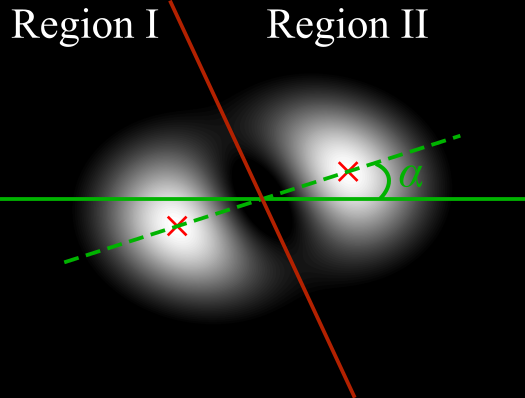}
	\caption{The red line is the nodal line, the green line is the horizontal, the center of mass of each region is demarcated by the red x and the green dashed line corresponds to the inclination of mode.}
	\label{fig:inclination}
\end{figure}

We developed an algorithm to calculate the mode inclination and visibility. First, we find the center of mass of the image and sum the intensity under a line passing through the center of mass which makes an angle $\eta$ with the horizontal, as shown in  Fig. \ref{fig:varrer}. The next step is to vary the $\eta$ angle between $0$ and $\pi$, so we can obtain a graph of the summed intensity versus $\eta\,$, an example of which is shown in Fig. \ref{fig:grafico3}. The orientation $\eta_{min}$ giving the minimum intensity defines the nodal line. Finally, two centers of mass are calculated, one on each side o the nodal line, and a line connecting them is traced. The mode inclination is determined by the orientation ($\alpha$) of this connecting line as described in Fig. \ref{fig:inclination}.  

The image visibility is defined as the ratio
\begin{equation}
v=\frac{I_{max}-I_{min}}{I_{max}+I_{min}},
\end{equation}
where $I_{max}$ and $I_{min}$ are, respectively, the maximum and minimum intensities obtained from the curve displayed in Fig. \ref{fig:grafico3}.

%

%\section*{References}

%\bibliography{bibliography}

\end{document}